\documentclass[a4paper,fleqn,usenatbib]{mnras}

\usepackage{mathptmx}
\usepackage[T1]{fontenc}
\usepackage{ae,aecompl}

\usepackage{graphicx}	% Including figure files
\usepackage{amsmath}	% Advanced maths commands
\usepackage{amssymb}	% Extra maths symbols

\usepackage{times}
\usepackage{multirow}
\usepackage[table]{xcolor}
\title[Radio follow-up of B0218+357]
  {Radio follow-up of the $\gamma$-ray flaring gravitational lens JVAS B0218+357}
  \author[C. Spingola et al.]{C. Spingola,$^{1,2,3}$\thanks{E-mail: spingola@astro.rug.nl} D. Dallacasa,$^{1,2}$ M. Orienti,$^{2}$ M. Giroletti,$^{2}$ J. P. McKean,$^{3,4}$ 
  \newauthor C. C. Cheung,$^{5}$  T. Hovatta,$^{6}$ S. Ciprini,$^{7,8}$ F. D'Ammando,$^{1,2}$ E. Falco,$^{9}$
  \newauthor S. Larsson,$^{10,11,12}$ W. Max-Moerbeck,$^{13}$ R. Ojha,$^{14,15,16}$ A. C. S. Readhead,$^{17}$ 
  \newauthor J. L. Richards,$^{18}$ J. Scargle$^{19}$\\
  $^1$ Dipartimento di Astronomia, Universit\`{a} di Bologna, via Ranzani 1, I-40127, Bologna, Italy\\
  $^2$ INAF Istituto di Radioastronomia, via Gobetti 101, I-40129, Bologna, Italy\\
  $^3$ Kapteyn Astronomical Institute, University of Groningen, P.O. Box 800, 9700 AV Groningen, The Netherlands\\
  $^4$ Netherlands Institute for Radio Astronomy (ASTRON), P.O. Box 2, 7990 AA Dwingeloo, The Netherlands\\
  $^5$ Space Science Division, Naval Research Laboratory, Washington, DC 20375-5352, USA\\
  $^6$ Aalto University Mets\"ahovi Radio Observatory, Mets\"ahovintie 114, FI-02540 Kylm\"al\"a, Finland\\
  $^7$ Agenzia Spaziale Italiana (ASI) Science Data Center, I-00133 Roma, Italy\\
  $^8$ Istituto Nazionale di Astrofisica - Osservatorio Astronomico di Roma, I-00040 Monte Porzio Catone (Roma), Italy\\
  $^9$ Harvard-Smithsonian Center for Astrophysics, Cambridge, MA 02138, USA\\
  $^{10}$ KTH Royal Institute of Technology, Departement of Physics, and the Oscar Klein Centre, AlbaNova, SE-106 91 Stockholm, Sweden\\
  $^{11}$ The Oskar Klein Centre for Cosmoparticle Physics, AlbaNova, SE-106 91 Stockholm, Sweden \\
  $^{12}$ Supported by the Royal Swedish Academy Crafoord Foundation\\
  $^{13}$ National Radio Astronomy Observatory, P.O. Box 0, Socorro, NM 87801, USA\\
  $^{14}$ NASA Goddard Space Flight Center, 8800 Greenbelt Rd, Greenbelt, MD 20771, USA\\
  $^{15}$ University of Maryland, Baltimore County, 1000 Hilltop Circle, Baltimore, MD 21250, USA\\
  $^{16}$ The Catholic University of America, 620 Michigan Ave NE, Washington, DC 20064, USA\\
  $^{17}$ Cahill Center for Astronomy and Astrophysics, California Institute of Technology, Pasadena CA, 91125, USA\\
  $^{18}$ Department of Physics and Astronomy, Purdue University, 525 Northwestern Ave, West Lafayette, IN 47907, USA\\  
  $^{19}$ Space Sciences Division, NASA Ames Research Center, Moffett Field, CA 94035-1000, USA\\
  }

\begin{document}

% These dates will be filled out by the publisher
\date{Accepted 2016 November 13. Received 2015 December 18; in original form 2015 October 30}
\pagerange{\pageref{firstpage}--\pageref{lastpage}}

% Enter the current year, for the copyright statements etc.
\pubyear{2015}

% Don't change these lines
\maketitle

\label{firstpage}

 \begin{abstract}
  We present results on multifrequency Very Long Baseline Array (VLBA)
  monitoring observations of the double-image gravitationally lensed 
  blazar JVAS B0218+357. Multi-epoch observations started less than one month 
  after the $\gamma$-ray flare detected in 2012 by the Large Area Telescope on board {\it Fermi}, 
  and spanned a 2-month interval. The radio light curves did not reveal any significant 
  flux density variability, suggesting that no clear correlation between 
  the high energy and low-energy emission is present. This behaviour was 
  confirmed also by the long-term Owens Valley Radio Observatory monitoring data at 15 GHz. 
  The milliarcsecond-scale resolution provided by the VLBA observations 
  allowed us to resolve the two images of the lensed blazar, which
  have a core-jet structure. No significant morphological variation 
  is found by the analysis of the multi-epoch data, suggesting that the 
  region responsible for the $\gamma$-ray variability is located 
  in the core of the AGN, which is opaque up to the highest observing frequency of 22 GHz.
 \end{abstract}

\begin{keywords}
 gravitational lensing: strong; quasars: individual: JVAS B0218+357.
\end{keywords}

\section{Introduction}

In the context of radio-loud Active Galactic Nuclei (AGN), when the
relativistic jet is closely aligned with
the line-of-sight, the emission is enhanced by Doppler boosting. 
These AGN represent the blazar population \citep{1979ApJ...232...34B}.
Their spectral energy distribution (SED) is characterized by two broad peaks: one at low energies due 
to synchrotron emission and the other at high energies produced by Inverse Compton scattering.   
Due to the broad range of energies covered by the blazar SED, it is clear that multi-wavelength 
studies are necessary for a proper understanding of their astrophysical properties.
Blazars are characterized by rapid flux density variability in all bands. 
Following the causality argument, the short-term variability observed during flaring episodes
indicates that the emission takes place in compact regions. 
Therefore, to investigate such small-scales a primary requirement is milliarcsecond-scale angular
resolution,  which can be achieved using Very Long Baseline Interferometry (VLBI) in the radio band.

A powerful approach to obtain accurate information
about the structure of blazars is strong gravitational lensing. 
As a consequence of lensing, the radiation from the background source is deflected on paths with different 
lengths and is magnified. In this case, the appearence of multiple images of the lensed blazar 
(which are distorted according to the mass distribution of the lens) is expected and any intrinsic
variability episode should be detected in the different images with a time-delay.

The blazar JVAS\footnote{Jodrell Bank VLA Astrometric Survey.} B0218+357 was identified as a lensing system by \citet{1992LNP...406..140P}
and is one of the brightest lensed objects in the Cosmic Lens All-Sky Survey (CLASS) sample (\citealt{2003MNRAS.341...13B},
\citealt{2003MNRAS.341....1M}).
This radio source consists of two compact images (A and B) 
separated by $\sim 335$ mas
and an Einstein ring with a similar size (\citealt{1992AJ....104.1320O}, \citealt{2001MNRAS.322..821B}).
This lensed blazar has a redshift $z=0.94$ 
\citep{2003ApJ...583...67C} and a frequency dependent radio structure \citep{2006A&A...447..515M}.
The lensing galaxy is a spiral galaxy at redshift $z=0.68$
(\citealt{1993MNRAS.263L..32B}, \citealt{2005MNRAS.357..124Y}).
The two images are resolved by VLBI
and show the core and inner jet of the lensed source
\citep{2003MNRAS.338..599B}. 
Given its variable radio emission, JVAS B0218+357 is an ideal target to measure
the time delay predicted by the lensing theory. 
The time delay between variations in the two images
was measured in the radio band and is $\sim 10$ days
(\citealt{1996IAUS..173...37C}, \citealt{1999MNRAS.304..349B}, \citealt{2000ApJ...545..578C}).

JVAS B0218+357 was observed with the \textsl{Fermi}-Large Area Telescope (LAT) (\citealt{2009ApJS..183...46A}). 
A bright $\gamma$-ray flare from JVAS B0218+357 was detected with the LAT in 2012 August \citep{2012ATel.4343....1C},
followed by the possible
gravitationally lensed echo \citep{2012ATel.4371....1G}.
The $\gamma$-ray delay was then measured to be $11.46\pm0.16$ days
which is one day greater than the previous radio measurements \citep{2014ApJ...782L..14C}.
This can be taken as evidence of the $\gamma$-ray and radio emitting regions
being in different parts of the background source (\citealt{2014ApJ...788..139B}, \citealt{2015arXiv151102891B}).
This source was detected again by \textsl{Fermi}-LAT in 2014 \citep{2014ATel.6316....1B}, while the echo was
seen by the Major Atmospheric Gamma Imaging Cherenkov (MAGIC) telescope \citep{2014ATel.6349....1M}.
The \textsl{Fermi}-LAT continuously monitors hundreds of sources.
So far, only one other gravitationally lensed blazar, PKS 1830$-$211, has been detected 
during a gamma-ray flare,
and the possible gravitationally lensed echo was 
observed from this system around 20 days later (\citealt{2015ApJ...799..143A}, \citealt{2011A&A...528L...3B},
\citealt{2015ApJ...809..100B}, \citealt{2015NatPh..11..664N}).

The connection between the radio and $\gamma$-ray emission in blazars has been investigated 
for a long time, and has received 
renewed attention during the \textsl{Fermi}-LAT era.
The site of the $\gamma$-ray emission in blazars is still a matter of debate;
the $\gamma$-ray radiation may originate
in the radio emitting region or closer to the central super-massive black hole (SMBH). 
If the $\gamma$-ray and radio emission is due to a shock propagating along the jet, the 
$\gamma$-ray flare may be followed by a radio counterpart and the time delay between the two flares
depends on the opacity at the longer wavelengths. 
Sometimes, the 
ejection of a superluminal jet component is observed.
To test this hypothesis, multi-band and multi-epoch radio observations with milliarcsecond resolution
of $\gamma$-ray blazars are necessary.
With this aim, after the $\gamma$-ray flare in 2012,
we performed a triggered monitoring campaign with the Very Long Baseline Array (VLBA) of JVAS B0218+357.

With its milliarcsecond resolution, the VLBA is a primary 
tool to resolve compact radio sources, like JVAS B0218+357, with the aim of finding a possible correlation 
of the radio emission (outbursts, new pc-scale components, etc.) with episodes of high $\gamma$-ray activity.
Moreover, the VLBA capabilities are also ideal to perform a study of the opacity variations
within the radio core.
The role of the VLBA for lensed sources is fundamental for studying their substructures 
and for determining the spatially resolved light curves, which can be 
compared with the $\gamma$-ray light curve.

In this paper we present results 
obtained from the multi-frequency VLBA monitoring campaign of JVAS B0218+357.
The paper is organized as follows. Section 2 describes the radio observations
and the data reduction, in Section 3 we report the results, while a discussion and summary
are presented in Sections 4 and 5, respectively.

Throughout this paper, we assume $H_0=71\; \mathrm{km\; s^{-1}\; Mpc^{-1}}$,
$\Omega_M=0.27$, $\Omega_{\Lambda}=0.73$ in a flat Universe. The spectral 
index $\alpha$ is defined as $S(\nu) \propto \nu^{-\alpha}$.

\section{Radio Data}

\subsection{VLBA observations}

We performed VLBA observations (BC214; PI: Cheung) at 2.3 GHz (\textit{S}-band), 8.4 GHz 
(\textit{X}-band) and 22 GHz (\textit{K}-band).
The data were recorded at 512 Mbps.
The observations were carried out between 2012 September 24 and 2012 November 25,
with a separation of a few days. Sixteen epochs with 1 h observations at
22 GHz were interspersed in four instances with 1 h integration at 2.3/8.4 GHz to probe the
spectral behavior.
The 2.3 and 8.4 GHz observations were simultaneous (Table \ref{observations}).

Scans on the target JVAS B0218+357 of 4 min each were interleaved by scans on the phase and bandpass
calibrator B0234+285.
The correlation was performed at the VLBA correlator in Socorro and
the data were processed with the Astronomical Image Processing Software (AIPS) package.
The a priori amplitude calibration was applied using measurements of the system temperature and
the antenna gains for each VLBA antenna.

Atmospheric opacity is not negligible at 22 GHz and all of the datasets at this frequency
were corrected for this effect.
The gain fluctuations found during the experiment were of the order of 7-8 per cent at 2.3/8.4 GHz 
and 10 per cent at 22 GHz, which we consider as a conservative estimate
of the absolute flux-density calibration error.
JVAS B0218+357 is strong enough to allow for fringe fitting.
Since the structure of the target is complex, a first-guess, point-like model of the source 
was used. 
Then we obtained a preliminary source model of the target by cleaning the fields
with the two images. This model was used for the fringe fitting in order to obtain a better determination
of the delay and rate of the phase variations.
\begin{table}
\centering
\small
\caption{Observational parameters of B0218+357. Column 1: Observing epoch; column 2: image; column 3: flux density at 2.3 GHz;
 column 4: flux density at 8.4 GHz; column 5: flux density at 22 GHz.}\label{observations}
\renewcommand\arraystretch{1.3}
 \begin{tabular}{c|c|ccc}
 \hline
  \textbf{Epoch} & \textbf{Image} & \textbf{$S_{\mathrm{2.3 GHz}}$} &  \textbf{$S_{\mathrm{8.4 GHz}}$} & \textbf{$S_{\mathrm{22 GHz}}$}\\
 &  &(mJy) & (mJy)  & (mJy)\\

\hline
 \multirow{2}[1]*{ 24/09/2012} &  A  & 805 $\pm$ 65 & 677 $\pm$ 48 & 499 $\pm$ 50 \\ \cline{2-5}
  &   B   &304 $\pm$ 22 & 227 $\pm$ 16 & 143 $\pm$ 11\\ 
 \hline
 
 \multirow{2}[1]*{ 27/09/2012} &  A  & & & 516$\pm$ 52\\ \cline{2-5}
    &   B   &  &  &142 $\pm$ 14\\ 
  \hline
  
 \multirow{2}[1]*{2/10/2012} &  A  & 807 $\pm$ 57& 643 $\pm$ 51 & 622 $\pm$ 62\\ \cline{2-5}
   &   B   &322 $\pm$ 23& 205 $\pm$ 16& 177 $\pm$ 18\\ 

  \hline
 \multirow{2}[1]*{5/10/2012} &  A  &  &  & 659$\pm$ 66\\ \cline{2-5}
   &   B   &   &  & 183$\pm$ 18\\ 

  \hline
 \multirow{2}[1]*{6/10/2012} &  A &  &  & 648$\pm$ 65\\ \cline{2-5}
&   B   &  & & 179$\pm$ 18\\ 
  \hline

 \multirow{2}[1]*{12/10/2012} &  A & &  &634 $\pm$ 63\\ \cline{2-5}
   &   B   &  & & 187$\pm$ 19\\ 
  \hline

 \multirow{2}[1]*{15/10/2012} &  A & &  & 632$\pm$ 63\\ \cline{2-5}
   &   B   &  &  &164 $\pm$ 16\\ 
  \hline

 \multirow{2}[1]*{19/10/2012} &  A & & & 628$\pm$ 63\\ \cline{2-5}
   &   B   &  & & 184$\pm$ 18\\ 
  \hline

 \multirow{2}[1]*{23/10/2012} &  A & 840$\pm$ 59& 690$\pm$ 55 & 618$\pm$ 62\\ \cline{2-5}
   &   B   & 315 $\pm$22 & 235$\pm$ 19& 178$\pm$ 18\\ 
  \hline

 \multirow{2}[1]*{26/10/2012} &  A & & & 613$\pm$ 61\\ \cline{2-5}
  &   B   &  & & 185$\pm$ 19\\ 
  \hline

 \multirow{2}[1]*{31/10/2012} &  A & & & 597 $\pm$60 \\ \cline{2-5}
   &   B   &  &  &167$\pm$ 17\\ 
  \hline

 \multirow{2}[1]*{5/11/2012} &  A &  &  & 652$\pm$ 65\\ \cline{2-5}
    &   B   &  &  & 171$\pm$17\\ 
  \hline

 \multirow{2}[1]*{13/11/2012} &  A &  &  & 652$\pm$ 65\\ \cline{2-5}
   &   B   &  &  &160 $\pm$ 16\\
  \hline

 \multirow{2}[1]*{15/11/2012} &  A &  &  &611 $\pm$ 61\\ \cline{2-5}
    &   B   &  & & 171$\pm$ 17\\ 
  \hline

 \multirow{2}[1]*{21/11/2012} &  A & & & 643$\pm$64\\ \cline{2-5}
   &   B   & & & 160$\pm$ 16 \\ 
  \hline

 \multirow{2}[1]*{25/11/2012} &  A & 847$\pm$ 60  & 652$\pm$52 & 593 $\pm$ 60\\ \cline{2-5}
    &   B   &303 $\pm$ 21 & 213 $\pm$ 21 & 173$\pm$ 17\\ 

 \hline
 \end{tabular}
 
\end{table}

\subsection{OVRO observations}

JVAS B0218+357 is part of an ongoing blazar monitoring programme at 15 GHz with
the OVRO 40-m radio telescope. This monitoring programme includes over 
1500 confirmed and candidate $\gamma$-ray-loud blazars above declination 
$-20^{\circ}$ (\citealt{2011ApJS..194...29R}). 
The sources in this programme are 
observed in total intensity twice per week with a 4 mJy (minimum) and 3 
per cent (typical) uncertainty. Observations are performed with a 
dual-beam (each with a 2.6 arcmin Full Width Half Maximum; FWHM) Dicke-switched system using the cold sky in 
the off-source beam as the reference. Additionally, the source is 
switched between beams to reduce atmospheric variations. The absolute 
flux density scale is calibrated using observations of 3C 286, adopting 
the flux density (3.44 Jy) from \citet{1977A&A....61...99B}. This results in 
about a 5 per cent absolute flux-density scale uncertainty, which is not reflected in 
the plotted errors.

\section{Results}
\subsection{VLBA images}

The final VLBA images were produced after a number of phase self-calibration iterations.
At the end of the process, we applied amplitude self-calibration to
remove residual systematic errors.

\begin{figure*}
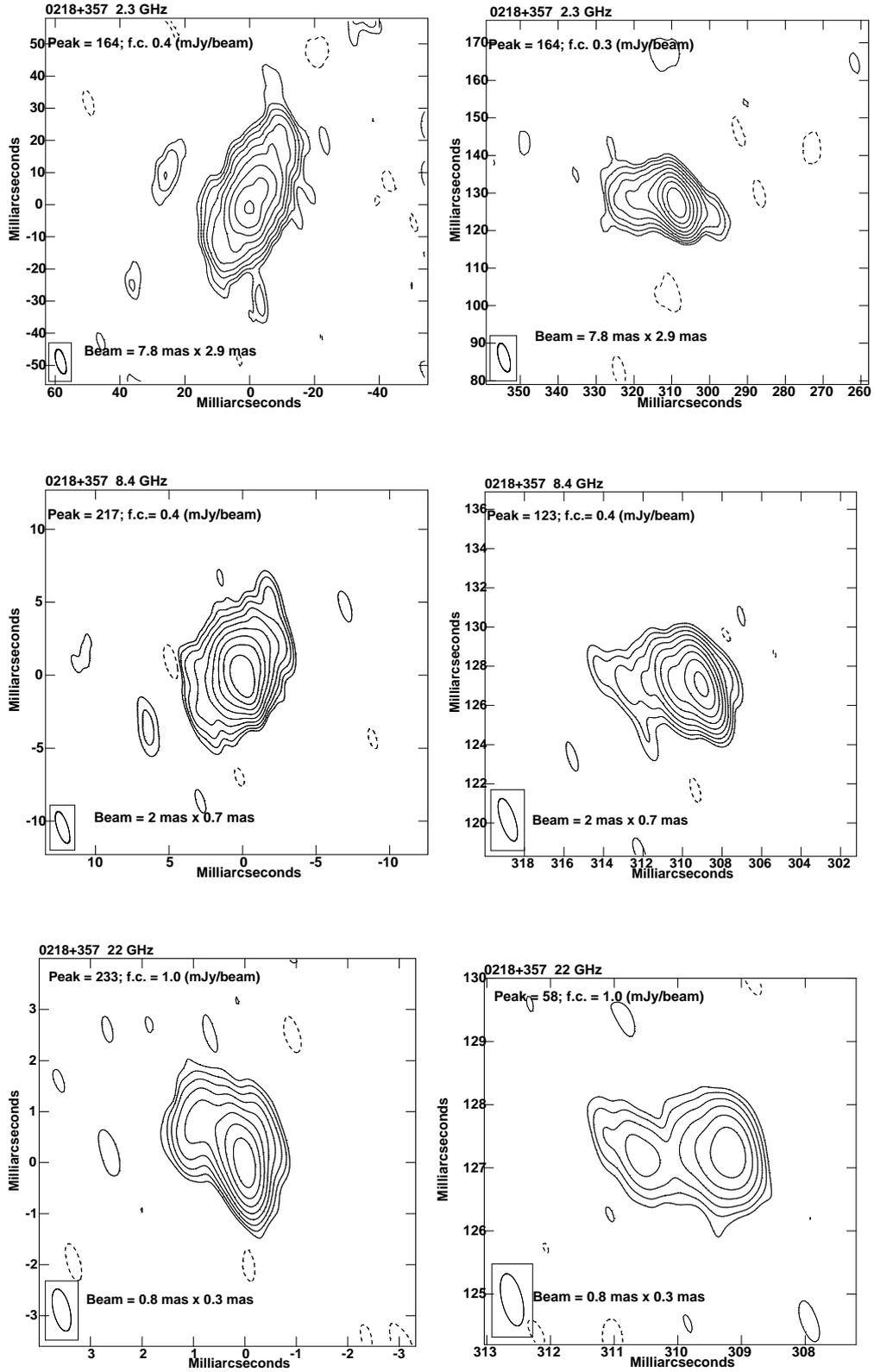

  \begin{minipage}{130mm}
  \includegraphics[scale=0.35]{fig1.ps}
  \includegraphics[scale=0.35]{fig2.ps}
  \includegraphics[scale=0.35]{fig3.ps}
  \includegraphics[scale=0.34]{fig4.ps}
  \includegraphics[scale=0.34]{fig5.ps}
  \includegraphics[scale=0.35]{fig6.ps}
\caption{Images of A (left) and B (right) at 2.3 GHz (top panel), 8.4 GHz (central panel) and 22 GHz (bottom panel).
The beam is plotted on the bottom left corner of each image. 
For each image, we provide the peak flux density in mJy $\mathrm{beam}^{-1}$, the first contour (f.c.) intensity 
(mJy $\mathrm{beam}^{-1}$), 
which is three times the off-source noise level measured on the image plane and the
size of the beam. Contour levels increase by a factor of 2.}\label{images}
  \end{minipage}
\end{figure*}

The final root-mean-square (rms) noise (1$\sigma$) measured on
the image plane is between 0.07 and 0.11 mJy $\mathrm{beam}^{-1}$ at 2.3 GHz,
between 0.08 and 0.10 mJy $\mathrm{beam}^{-1}$ at 8.4 GHz
and between 0.08 and 0.21 mJy $\mathrm{beam}^{-1}$ at 22 GHz.
The restoring beam at 2.3, 8.4 and 22 GHz is 7.8 mas $\times$ 2.9 mas,  2 mas $\times$ 0.7 mas, and
0.8 mas $\times$ 0.3 mas, respectively.
The total flux density of each image, as well as the extended 
emission, was measured by using an 
aperture integration over a selected region on the image plane.
The total flux density of the two images, A and B,
for all the observing epochs, is reported in Table \ref{observations}.
Images of A and B at 2.3, 8.4 and 22 GHz are shown in Fig. \ref{images}.

\begin{figure*}
 \begin{minipage}{150mm}
 \centering
    \includegraphics[scale=0.4]{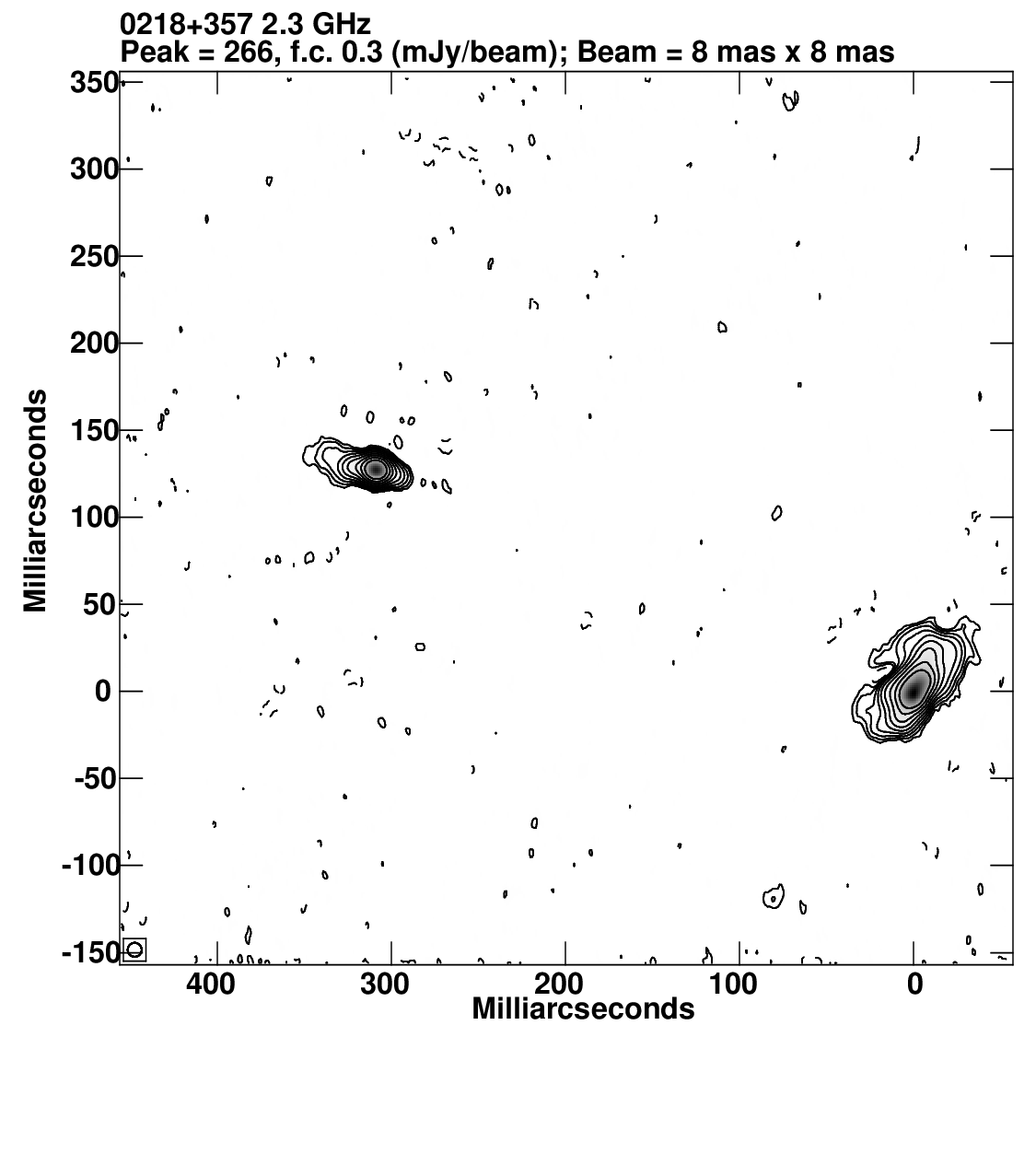}
    \caption{\small Stacked image of B0218+357 at 2.3 GHz. The beam is plotted on the bottom left corner. 
We provide the peak flux density in mJy $\mathrm{beam}^{-1}$, the first contour (f.c.) intensity (mJy $\mathrm{beam}^{-1}$), 
which is three times the off-source noise level measured on the image plane and the
size of the beam. Contour levels increase by a factor of 2.}\label{stacked_s}
  \end{minipage}
\end{figure*}

\begin{figure*}
 \begin{minipage}{150mm}

    \includegraphics[scale=0.4]{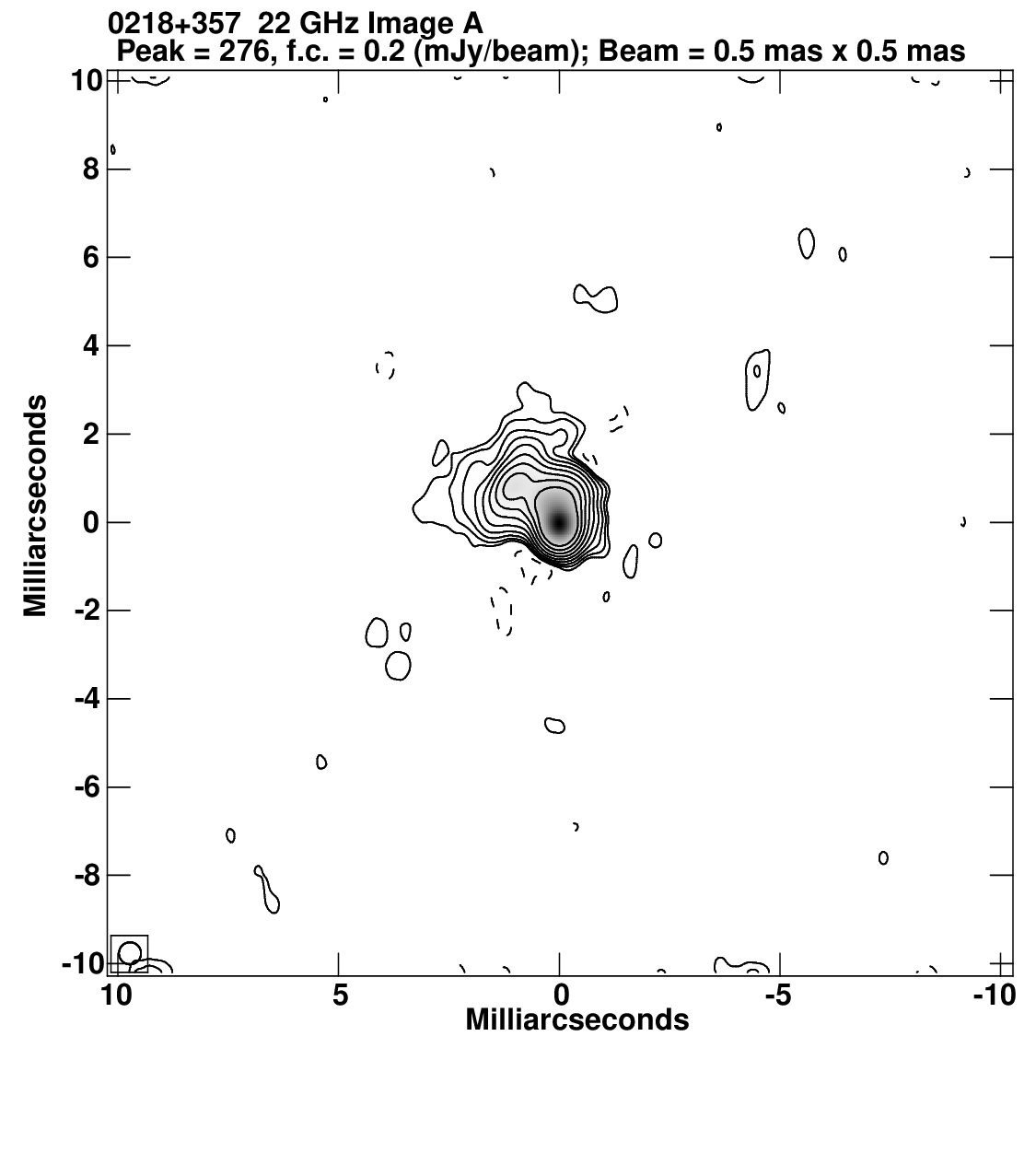}
    \includegraphics[scale=0.4]{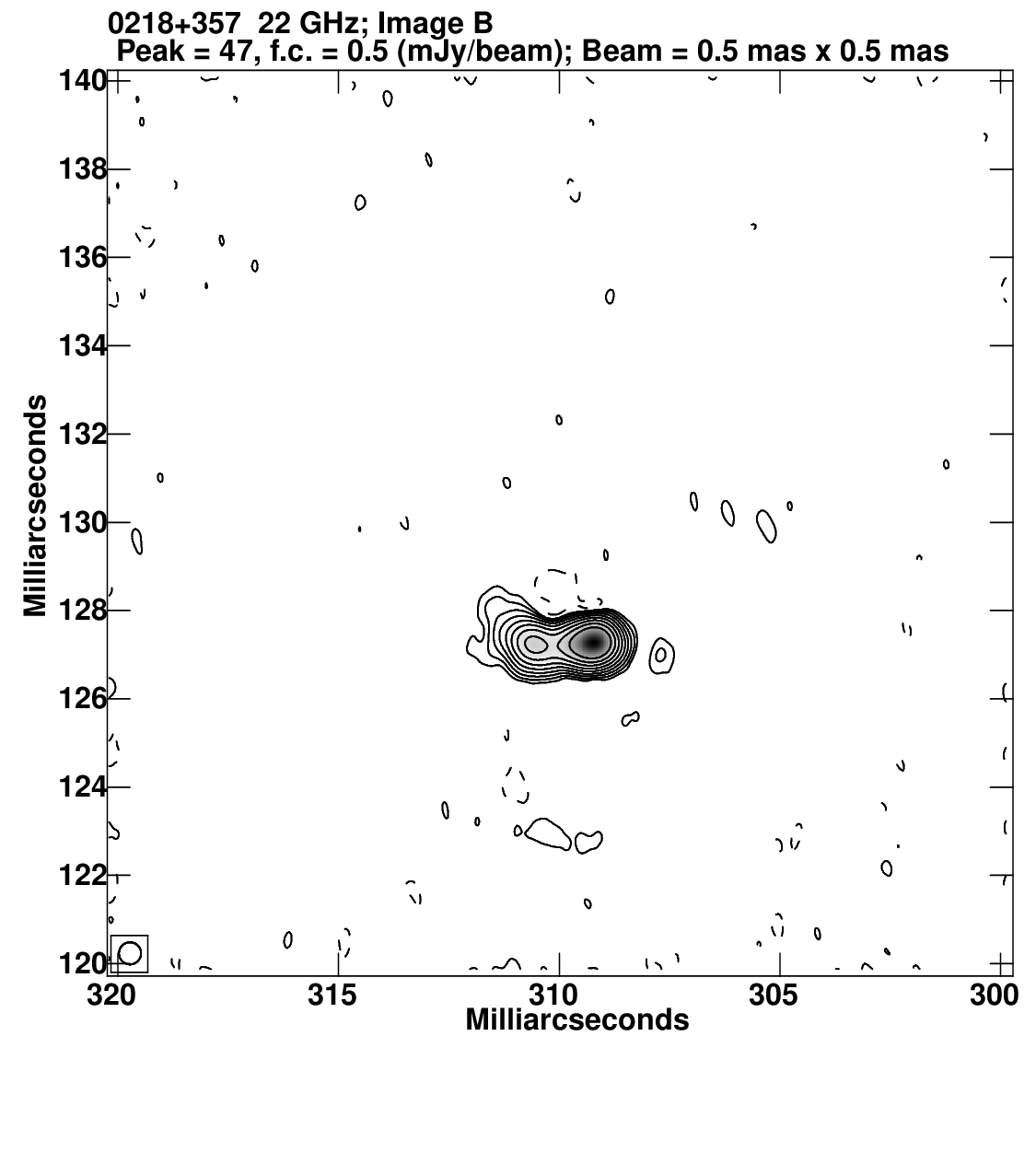}
    \caption{\small Stacked images of B0218+357 at 22 GHz. 
    \textsl{Left:} Stacked image of A. \textsl{Right:} Stacked image of B. 
    The beam is plotted on the bottom left corner. 
    We provide the peak flux density in mJy $\mathrm{beam}^{-1}$, the first contour (f.c.) intensity (mJy $\mathrm{beam}^{-1}$), 
    which is three times the off-source noise level measured on the image plane and the
    size of the beam. Contour levels increase by a factor of 2.}\label{stacked_k}
 \end{minipage}
\end{figure*}

To improve the signal-to-noise ratio, at 2.3 and 22 GHz we produced a 
high-sensitivity image by stacking all of the images at the various epochs. 
Each image was reconstructed with the same circular beam (i.e. 8 mas at 
2.3 GHz, and 0.5 mas at 22 GHz), and the alignment was checked by comparing the 
position of the peak of the optically-thin jet base in image B.
The images are shown in Figs. \ref{stacked_s} and \ref{stacked_k}.

The flux density and size 
of the compact sub-structures of each image were determined by 
a Gaussian fit on the image plane. The results are reported in Tables \ref{phys_param_A} and \ref{phys_param_B}.
The flux density errors are mainly due to the uncertainty on the absolute flux 
density scale (see Section 2.1), which dominates over the image rms.

\subsection{VLBA morphology: parsec-scale structure}
The high angular resolution of the VLBA observations allows us to
resolve both images A and B, and their sub-components,
which are likely the core and the jet.
At 22 GHz we achieve the best angular resolution, 
and the core-jet morphology of the source is well resolved (Fig. \ref{images}, bottom panel).
Image B does not show significant image distortion, although 
the structure of image A is rather complex,
as a result of the gravitational lensing and possibly scatter-broadening (\citealt{2003MNRAS.338..599B},
\citealt{2007A&A...465..405M}).
The overall position angle (p.a.) rotates from 2.3 to 22 GHz
as a combined effect of angular resolution and spectral index (Fig. \ref{images}).

In the stacked images at low resolution (2.3 GHz), it is possible to recognize the core-jet structure in image B, 
while image A is still too complex to discern the core from the jet.
Its emission is dominated by the diffuse component (Fig. \ref{stacked_s}). 
On the other hand, in the stacked images at 22 GHz it is possible to clearly recognize 
the typical core-jet structure in both of the images, although the source p.a. is different
due to scatter-broadening (Fig. \ref{stacked_k}).

 In order to quantify any possible variation in the jet of JVAS B0218+357, 
 we fit the self-calibrated
 images at 22 GHz with elliptical Gaussian components.
 We used these images since they have the best resolution among the frequencies studied here
 and the fit to the components would be more accurate. Moreover
 the 22 GHz observations have a dense time sampling (16 epochs). Therefore, they are suitable
 to study the variability of the target.
 We could find only marginal variations in the fit parameters, mostly arising from differences in the \textit{uv}-coverage, 
 rather than being intrinsic to the source. 
 Images of the fit residuals were analysed to evaluate possible additional components and to make sure that
 the fit parameters were a good representation of the source parameters.
 Therefore, the Gaussians components are an adequate depiction of the core and the jet, and they do not include any
 new knot along the jet.
  
 To obtain basic physical information of the target source, such as the peak intensity and its position, the flux density and the
 deconvolved size of the elliptical Gaussian regions, more attention has been paid to the stacked images.
 This is because they have better
 sensitivity, and reduce the artifacts coming from the limited uv-coverage. This contributes to enhancing the sampling of the
 visibility function that was eventually found to be non variable during the whole time range spanned by the VLBA observation.
 We fit image A with 4 Gaussian components and image B with 3 Gaussian components, and the 
 values are reported in Tables \ref{phys_param_A} and \ref{phys_param_B}. 
 We used the same nomenclature as
 \cite{1995MNRAS.274L...5P}: components A1 and B1 are interpreted as the core of JVAS B0218+357,
 A2, A3, A4, B2 and B3 are likely jet components.

\subsection{Light curves}

The VLBA flux density monitoring provides an important tool
for the analysis of the variability in this lensed blazar.
The 2.3/8.4 GHz observations were carried out to study the flux density 
variability of the core opacity rather than the total flux density variability. 
That is the reason why there are only four epochs at these frequencies.
A better time sampling was available at 22 GHz.
At this frequency the opacity should be less effective than 
at lower frequencies, and changes in the flux density may occur on 
shorter time scales.

\begin{figure*}
 \begin{minipage}{180mm}
  \centering
  \includegraphics[scale=0.4]{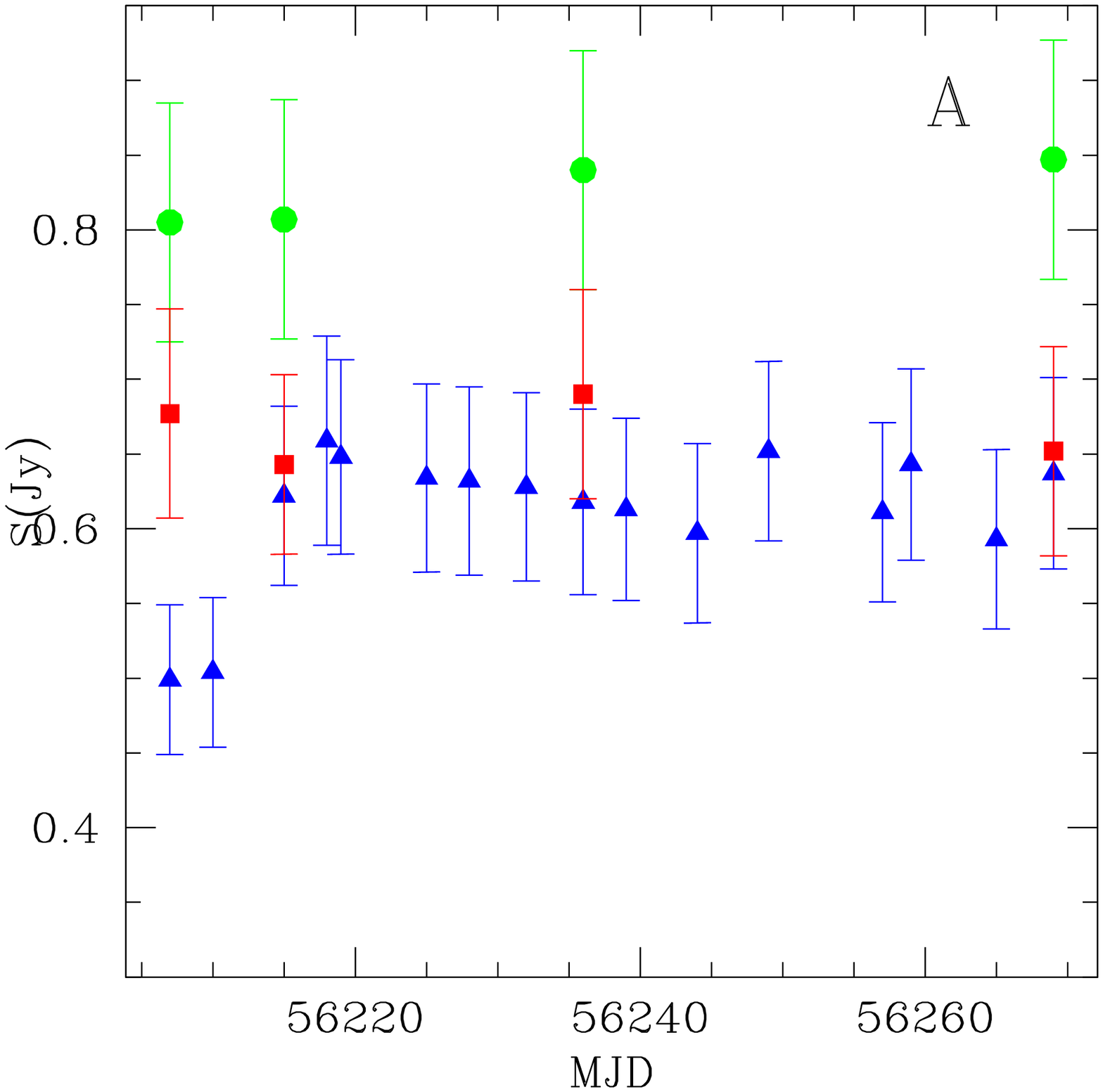}
  \includegraphics[scale=0.4]{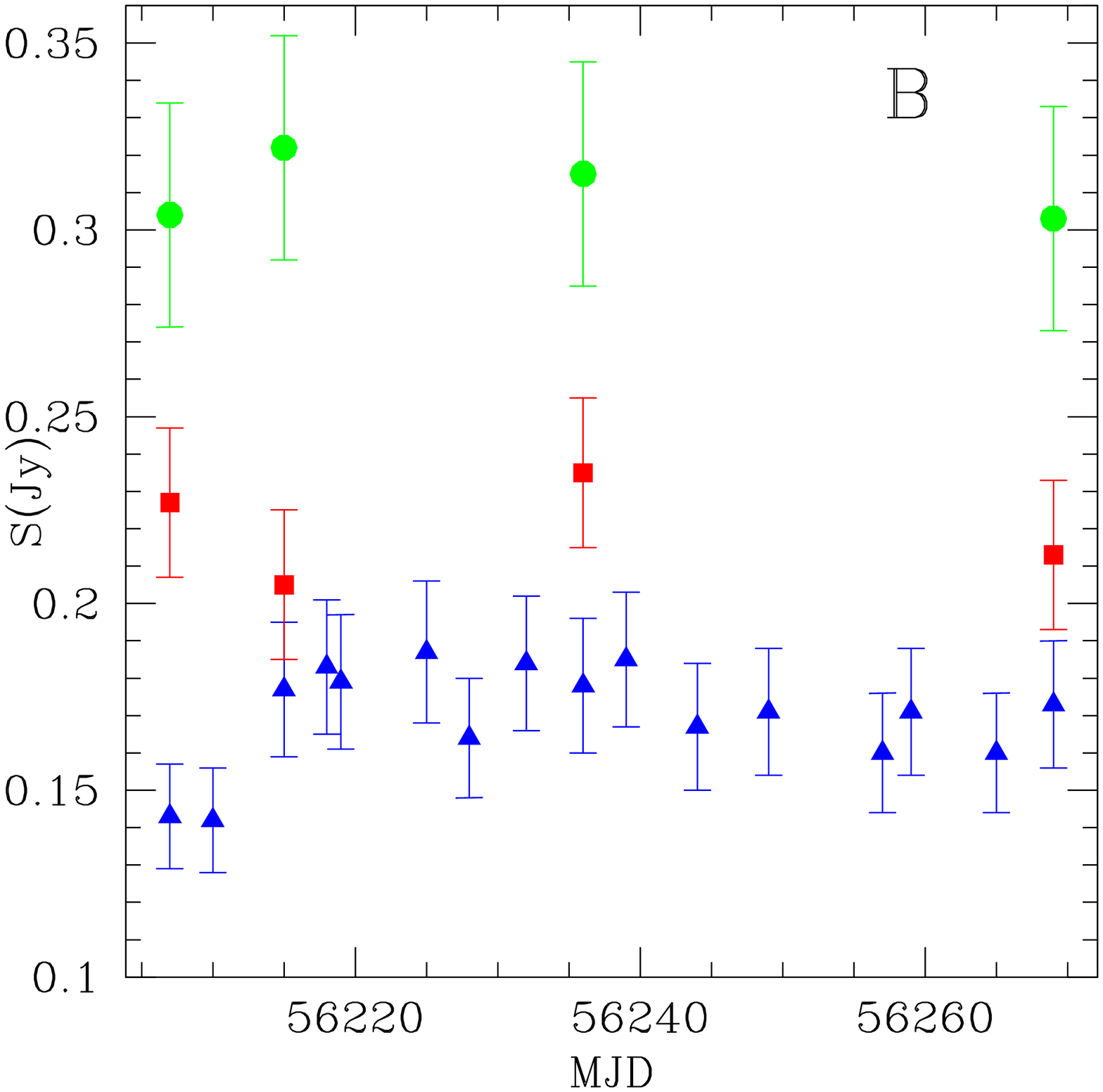}
  \caption{\small The light curves of images A (left) and  B (right) in the three bands. 
  Green circles represent 2.3 GHz data, red squares are the 8.4 GHz data and blue triangles are the 22 GHz data.}\label{lightcurves}
 \end{minipage}
\end{figure*}

Within the large errors of the absolute flux-density calibration 
the light curves show that the flux density is almost constant for all of 
the period of observations in the three bands, 
as shown in Fig. \ref{lightcurves}.
However, at 22 GHz the first and the second epoch data provided a 
flux density about 1$\sigma$ below the average flux density measured in the other experiments.
Such a difference is unlikely to be related to an intrinsic source variability, 
but it might be caused by amplitude calibration.
In fact, images A and B appear weaker simultaneously, without any signature of a time-delay, expected
in the case of an intrinsic flux density variation of a lensed source.
We reiterate the light curves are consistent with no variability,
due to the large uncertainty on the absolute flux density calibration from epoch to epoch.

 \begin{figure}
 %\begin{minipage}{180mm}
 \centering
   \includegraphics[scale=0.4]{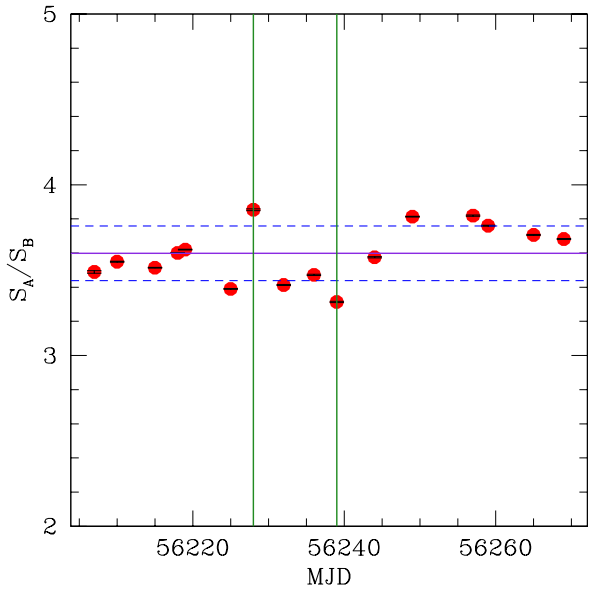}
  \caption{\small Flux ratio of A and B as a function of time at 22 GHz.
  The error bars represent the uncorrelated error, which is typically of the order of $\sim 0.002$ per cent. 
  The vertical green solid lines define the maximum and the minimum of the light curve 
  and are separated by $\sim$ 11 days. 
 The solid violet line represents the mean value of the magnification ratio, which is $3.59 \pm 0.16$.
  The 1$\sigma$ uncertainties are traced as dashed blue lines.}\label{ratio_time}
 %\end{minipage}
\end{figure} 

\begin{figure*}
 \begin{minipage}{180mm}
 \centering
  \includegraphics[scale=0.4]{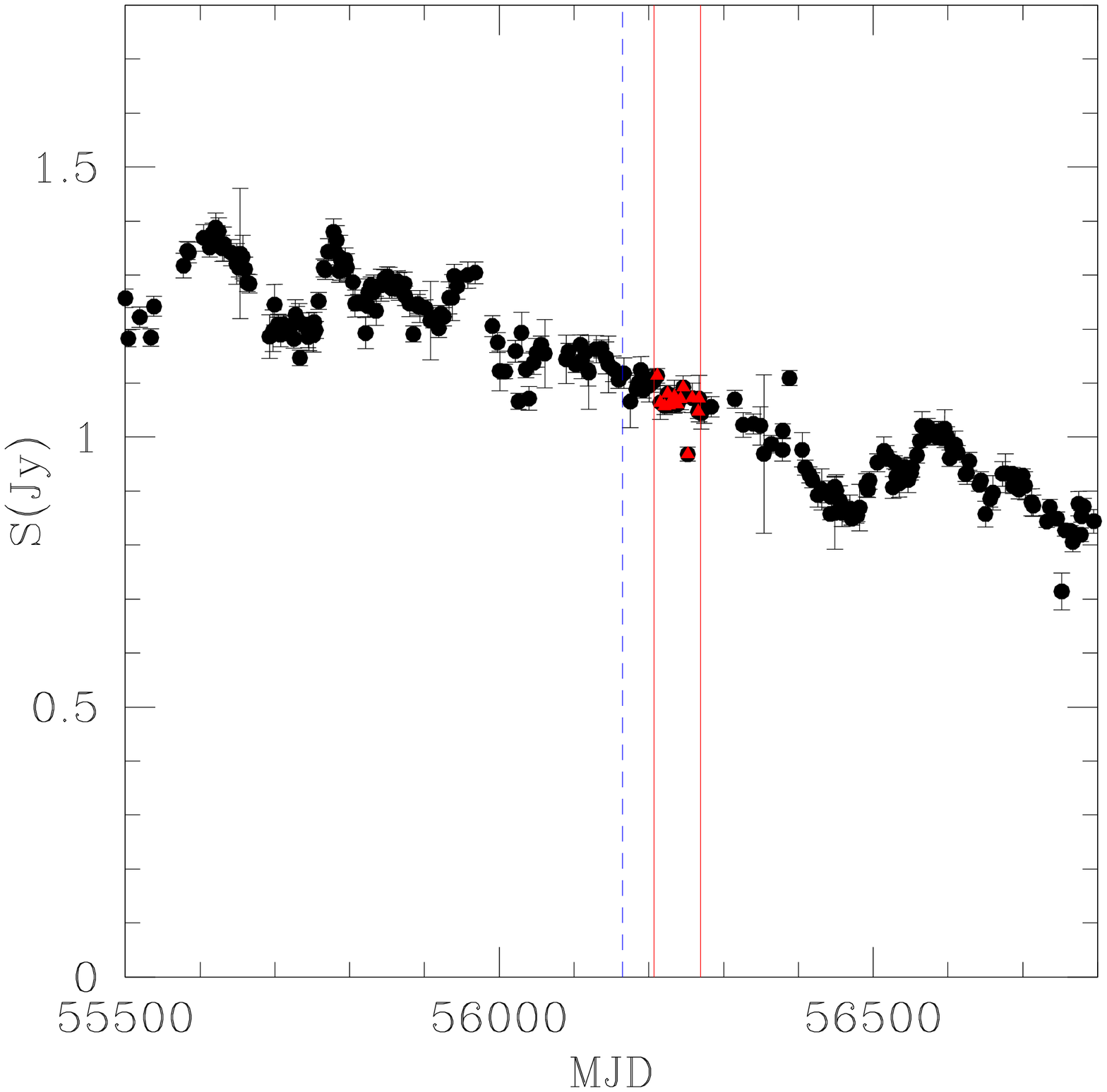}
  \includegraphics[scale=0.4]{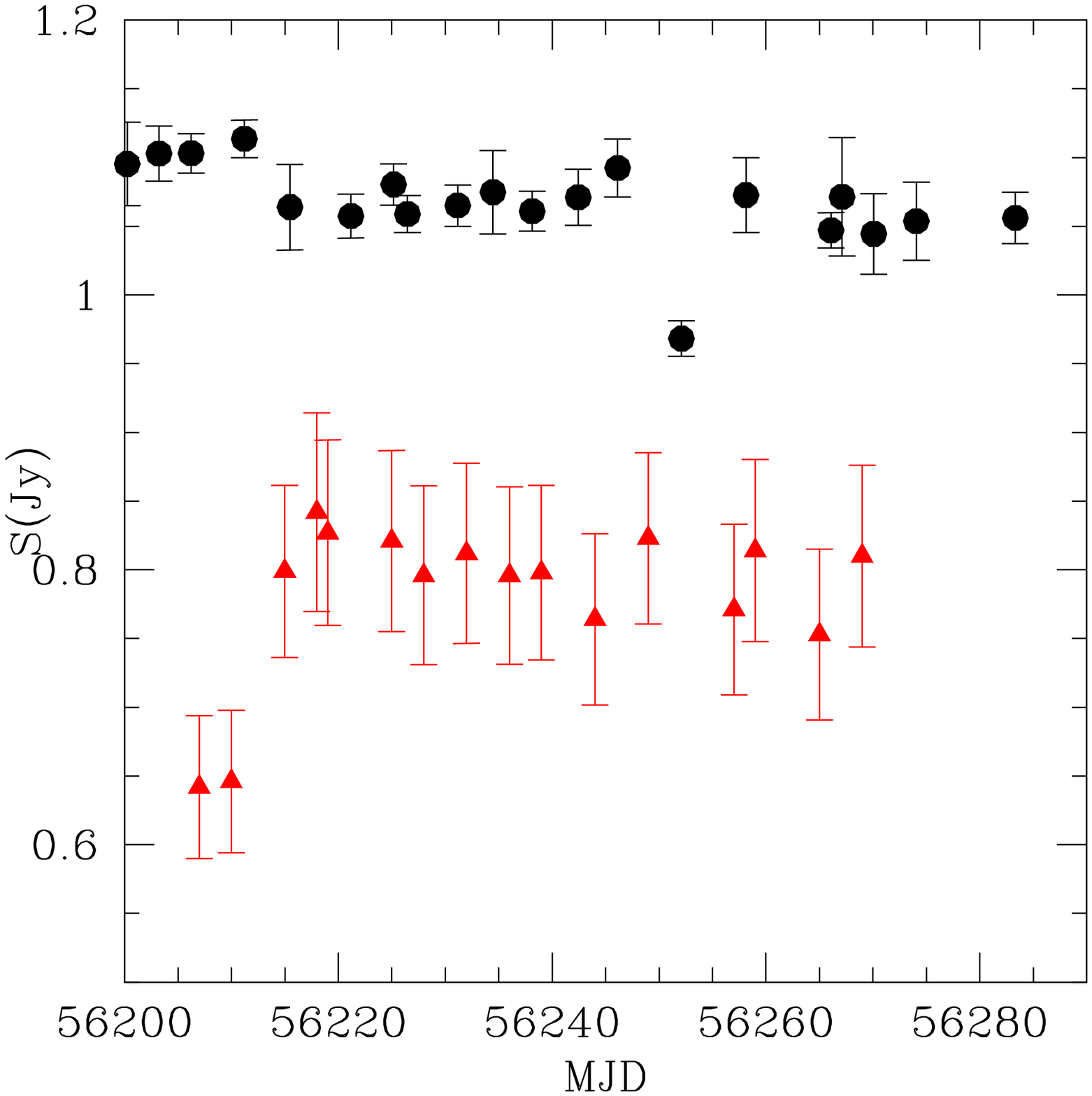}
  \caption{\small\textsl{Left:} The OVRO light curve of the target; two vertical red solid lines
  and red triangles indicate the period of VLBA observations,
  the dashed blue line indicates the \textsl{Fermi}-LAT detection of the flare. 
  \textsl{Right:} Zoom of the OVRO light curve of B0218+357 (black circles) during the period of VLBA observations 
  (MJD 556207 - MJD 56269). The 22 GHz light curve is represented by red triangles.}\label{ovro}
 \end{minipage}
\end{figure*}

However, since the flux density ratio is not affected by errors in the absolute flux calibration, 
we analysed the light curve of the flux density ratio (Fig. \ref{ratio_time}).
This is a standard method for looking for the small scale variablity within gravitational lenses,
because it avoids the systematic errors associated with the absolute flux calibration
(e.g. \citealt{2003ApJ...595..712K}).
This is not the preferred way for measuring the time delay, but it is a useful method in order to have an indication of it.
Moreover, if there is not any variability, the flux density ratio will be constant.
Indeed there was variability during this period of observations and it
can be associated with the intrinsic variability of the blazar
rather than the $\gamma$-ray flare.
Looking at the flux density ratio as a function of time  (Fig. \ref{ratio_time}),
between the first and the seventh epochs there is a rise of $\sim10$ per cent, 
which can be attributed to the increasing flux density of image A. 
There is evidence of a drop in flux ratio between the seventh and the tenth epochs
caused by the delayed enlightenment of image B.
This behaviour is expected since the source JVAS B0218+357 is intrinsically variable and the two 
images vary in the order A$\rightarrow$B. 
Given the scatter, the average magnification ratio is $3.59\pm0.16$.
Moreover, the time between the maximum and the minimum of the flux ratio change is 11 days
and the typical percentage of the uncorrelated errors is $\sim 0.002$ per cent.

As a further check, the VLBA light curves were compared with the light curve at 15 GHz
obtained by the OVRO data.
The OVRO single-dish has a resolution of 157 arcsec ($\sim 2.6$ arcmin), and, therefore,
it cannot resolve the source into images A and B.
Hence, we decided to compare the sum of the flux densities of component A and B with 
the OVRO flux density measurements (Fig. \ref{ovro}, right panel).
The OVRO light curve turned out to be in good agreement with our results 
and JVAS B0218+357 did not show any significant small time-scale
variability during 2012, with an exception of a longer decrease ($\sim 30$ per cent) 
of the flux density from March 2012 (Fig. \ref{ovro}, left panel).
However, between July and March 2013 a flux density enhancement was detected by OVRO.
This flux density increase might be related to both the $\gamma$-ray flare 
and the intrinsic flux density variablity of the source.

The sum of the flux density of images A and B at 22 GHz is lower than the OVRO measurements.
Since the observing frequency of OVRO is 15 GHz, this drop in the flux density 
may be assigned to the Einstein ring, which is not detected on the mas-scales of our
VLBA imaging. In fact the flux density of 
the Einstein ring was found to be $\sim 126$ mJy at 15 GHz \citep{1993MNRAS.261..435P}.

\section{Discussion}

The gravitationally lensed blazar JVAS B0218+357 is a variable source
at all wavelengths from $\gamma$-rays to the radio band. 
Multi-epoch high-resolution observations are a fundamental tool for
investigating flux density and morphological variability in blazars.
Some $\gamma$-ray flares in blazars are likely associated with
the appearance of new superluminal jet knots moving downstream the jet \citep{1985ApJ...298..114M}.
Kinematic model fitting of VLBI data indicated that the ejection of these knots occurs
close in time with a $\gamma$-ray flare 
(\citealt{2001ApJS..134..181J}, \citealt{2013ApJ...773..147J}).
If the knots are the observable manifestation of shocks, 
they can form near the base of the jet, causing a substantial increase in
both velocity and pressure of the flow. 
However, the location of the $\gamma$-ray and radio emission is still controversial. 

In 2012 August, a $\gamma$-ray flare was observed from JVAS B0218+357 followed by the lensed echo with an 11 day 
time-delay. The analysis of the radio light curve did not reveal any radio flare. The flux density
of the target was constant after the $\gamma$-ray flare without showing any increase, which
could have been related to the $\gamma$-ray flare. 
This may be explained either by assuming that the $\gamma$-ray flare originated in a
medium that is optically thick to the radio wavelengths or
the $\gamma$-ray emission and the radio emission are not correlated.
\cite{2014ApJ...782L..14C} estimated the size of the $\gamma$-ray emitting
region to be less than $6\times 10^{14}\;\mathrm{cm}$. 
This value is smaller than the previous radio constraint of the core region \citep{2007A&A...465..405M}. 
Another constraint on the $\gamma$-ray
emitting region was obtained by microlensing measurements \citep{2015arXiv150701092V},
which is consistent with the value obtained by 
\cite{2014ApJ...782L..14C} under the assumption of a Lorentz factor of $\sim 1$.
Our hypothesis is that this compact $\gamma$-ray emitting region is opaque to 
the radio band and, therefore, we were not able to detect the radio counterpart of the $\gamma$-ray flare.
However, there is an apparent difference between the $\gamma$-ray delay and radio delay of 1 day.
This difference can be attributed to a physical offset between the $\gamma$-ray
and the radio emitting regions of $\sim10$ mas ($\sim 70$ pc projected), as postulated by \cite{2014ApJ...782L..14C}.
Following this hypothesis, \citet{2015arXiv151102891B} 
have confirmed that there should be a physical displacement between the radio core and the
$\gamma$-ray emitting region.

JVAS B0218+357 is not a unique case of a $\gamma$-ray only flaring blazar.
For example, PKS 1830$-$211 experienced a bright $\gamma$-ray flare in 2010 October 
without showing any counterpart at lower
frequencies (NIR-optical and X-ray). \cite{2011ApJ...736L..30D} explain this behaviour in terms of a ``steady''
electron population. 
Another case is represented by the blazar PKS 1510$-$089. 
The $\gamma$-ray flare observed in 2009 March is not correlated with the radio and X-ray emission,
instead it seems correlated with the optical emission \citep{2010ApJ...721.1425A}.
Moreover, for most sources, \cite{2014MNRAS.445..428M} did not detect significant correlation between
the $\gamma$-ray and radio variations in the long-term analysis of the 
brightest blazars detected by the \textsl{Fermi}-LAT and monitored by OVRO, indicating a
difficult multi-wavelength connection for the majority of the sources.
On the other hand, performing a stacking analysis,
\cite{2014MNRAS.441.1899F} found a highly significant 
multiband radio and $\gamma$-ray correlation for a sample of \textsl{Fermi}-LAT bright 
blazars, leaving open the debate on the origin of the radio and $\gamma$-ray emission.

As previously mentioned, the 
detection of strong $\gamma$-ray flares from blazar objects may occur close 
in time with the ejection of a new superluminal jet component. No 
evidence of a superluminal component is found in JVAS B0218+357 by the 
comparison of the multi-epoch images at 22 GHz. This may be related to 
the short time range spanned by our VLBA observations. We set an upper 
limit on the apparent separation velocity ($v_{app}$) using
$v_{app}= D_A \times (1+z) \times \vartheta / \Delta t_{obs}$,
where $D_A$ is the luminosity distance in cm, $\vartheta$ is the minor axis of the beam at 22 GHz in radians, 
$z$ is the redshift and $\Delta t_{obs}$ is the period of the observation in seconds.
Therefore, the $v_{app}$ expected at 22 GHz is about $\sim 90c$, which is unlikely and is not
observed even in the most extreme blazars
(\citealt{2001ApJS..134..181J}, \citealt{2009AJ....138.1874L}, \citealt{2013AJ....146..120L}, \citealt{2013MNRAS.428.2418O}).
For this reason we conclude that no new knots could have been detected with 
our VLBA campaign.
The brightness temperature $T_B$ estimated in both the stacked images is about of $T_B\gtrsim(4-9) \times 10^{10}$ K 
for the core component and of $T_B\gtrsim(1-6)\times 10^{9}$ K for the jet 
components.  Given that the measured values of $T_B$ are lower limits, 
it is likely that they are affected by beaming effects \citep{1994ApJ...426...51R}.

\section{Summary}
We have presented results on a multi-frequency VLBA monitoring campaign of the 
$\gamma$-ray flaring gravitationally lensed blazar JVAS B0218+357. The 
observations were spread over a two-month period and started roughly one 
month after the $\gamma$-ray flaring activity detected by \textsl{Fermi}-LAT on 2012 August.
Although the source was very active in the $\gamma$-rays, 
it did not show any significant variability in the radio band, 
precluding us from investigating any possible connection between the 
high-energy and low-energy emission. On time-scales longer than those 
spanned by our VLBA observations, the OVRO light curve at 15 GHz 
indicates some variability, but the change in the flux density is within 
10 per cent, i.e. much smaller than what is found at high energies. The 
lack of a radio flare may suggest that the region responsible for the 
$\gamma$-ray activity is compact and located close to the innermost part of 
the AGN, which is opaque to the radio wavelengths. The high angular 
resolution of our VLBA data allowed us to resolve the source images into 
subcomponents and determining their core-jet structure. Some distortion 
effects are present in component A, where at 2.3 and 8.4 GHz the core-jet 
structure is hidden by diffuse emission.
No evidence of a change in the source morphology was found by comparing 
the multi-epoch images at 22 GHz, supporting the steady behaviour of the 
source in the radio band.

\begin{table*}
 \begin{minipage}{180mm}
\centering
\caption{Observational parameters of image A at 22 GHz. Column 1: Gaussian component; Column 2: peak intensity; Column 3: total flux
density; Column 4: deconvolved major axis; Column 5: deconvolved minor axis; Column 6: position angle of the major axis;
Column 7: position relative to component A1.}\label{phys_param_A}
\begin{tabular}{ccccccc}
\hline
 \multicolumn{7}{c}{\textbf{Image A}}\\
\hline
\textsl{Comp.}& \textsl{$S_P$}& \textsl{$S_t$} & \textsl{$\vartheta_{maj}$}  & \textsl{$\vartheta_{min}$} & \textsl{p.a.} & \textsl{Pos.}\\
 &  (mJy beam$^{-1}$)& (mJy) & (mas)  & (mas) & (deg) & (mas)\\
\hline
A1 & $259 \pm 26$ & $333 \pm 33$ & $0.355$ & $0.158$ & $4$  & (0.0, 0.0)    \\
A2 & $65  \pm 7$  & $132 \pm 13$ & $0.523$ & $0.496$ & $142$& (0.01, 0.01)  \\
A3 & $43  \pm 4$  & $59  \pm 6$  & $0.438$ & $0.125$ & $155$& (0.02, 0.31)  \\
A4 & $33  \pm 3$  & $84  \pm 8$  & $0.857$ & $0.395$ & $143$& (0.01, 1.50)  \\
 \hline

\end{tabular}
\end{minipage}
\end{table*}

 \begin{table*}
   \begin{minipage}{180mm}

 \centering
\caption{Observational parameters of image B at 22 GHz. Column 1: Gaussian component; Column 2: peak intensity; Column 3: total flux density;
Column 4: deconvolved major axis; Column 5: deconvolved minor axis; Column 6: position angle of the major axis; Column 7: position relative 
to component A1.}\label{phys_param_B}
    \begin{tabular}{ccccccc}
\hline
 \multicolumn{7}{c}{\textbf{Image B}}\\
\hline
\textsl{Comp.}& \textsl{$S_P$}& \textsl{$S_t$} & \textsl{$\vartheta_{maj}$}  & \textsl{$\vartheta_{min}$} & \textsl{p.a.} & \textsl{Pos.}\\
 &  (mJy beam$^{-1}$)& (mJy) & (mas)  & (mas) & (deg) & (mas)\\
\hline
B1 & $69 \pm  7 $ & $78\pm 7$ & $0.233$ & $0.132$ & $99$ & (127.21, 309.20) \\
B2 & $35 \pm 4$   & $62\pm 6$ & $0.655$ & $0.191$ & $89$ & (127.01, 310.91) \\
B3 & $17\pm 2$    & $29\pm 3$ & $0.534$ & $0.264$ & $66$ & (127.02, 310.51) \\
 \hline
    \end{tabular}
  \end{minipage}
\end{table*}

\section*{Acknowledgments}
The VLBA is operated by the US National Radio Astronomy Observatory 
which is a facility of the National Science Foundation operated under a 
cooperative agreement by Associated University, Inc., under contract 
with the National Science
Foundation.
This research has made use of the NASA/IPAC Extragalactic
Database (NED), which is operated by the Jet Propulsion Laboratory,
California Institute of Technology, under contract with the National
Aeronautics and Space Administration.

C.C.C. was supported at NRL by NASA Guest Investigator program 13-FERMI13-0009.
We thank the NRAO Schedsoc for approving our request and Mark Claussen for facilitating 
the VLBA scheduling.
The OVRO 40-m monitoring program is
supported in part by NASA grants NNX08AW31G
and NNX11A043G, and NSF grants AST-0808050
and AST-1109911.

This research was funded in part by NASA through Fermi Guest
Investigator grants NNH09ZDA001N, NNH10ZDA001N, NNH12ZDA001N,
NNH13ZDA001N-FERMI. This research was supported by an appointment to
the NASA Postdoctoral Program at the Goddard Space Flight Center,
administered by Oak Ridge Associated Universities through a contract
with NASA.

\label{lastpage}

\end{document}